# Decentralized Overlay for Federation of Enterprise Clouds

*Rajiv Ranjan and Rajkumar Buyya*
***Gri**d Computing and **D**istributed **S**ystems (GRIDS) Laboratory*
*Department of Computer Science and Software Engineering*
*The University of Melbourne, Australia*
Email*: {rranjan, raj}@csse.unimelb.edu.au*

## Abstract

This chapter describes Aneka-Federation, a decentralized and distributed system that combines enterprise Clouds, overlay networking, and structured peer-to-peer techniques to create scalable wide-area networking of compute nodes for high-throughput computing. The Aneka-Federation integrates numerous small scale Aneka Enterprise Cloud services and nodes that are distributed over multiple control and enterprise domains as parts of a single coordinated resource leasing abstraction. The system is designed with the aim of making distributed enterprise Cloud resource integration and application programming flexible, efficient, and scalable. The system is engineered such that it: enables seamless integration of existing Aneka Enterprise Clouds as part of single wide-area resource leasing federation; self-organizes the system components based on a structured peer-to-peer routing methodology; and presents end-users with a distributed application composition environment that can support variety of programming and execution models. This chapter describes the design and implementation of a novel, extensible and decentralized peer-to-peer technique that helps to discover, connect and provision the services of Aneka Enterprise Clouds among the users who can use different programming models to compose their applications. Evaluations of the system with applications that are programmed using the Task and Thread execution models on top of an overlay of Aneka Enterprise Clouds have been described here.

## 1. Introduction

Wide-area overlays of enterprise Grids [4][11][14][18] and Clouds [25][26][27][32] are an appealing platform for the creation of high-throughput computing resource pools and cross-domain virtual organizations. An enterprise Cloud[1] is a type of computing infrastructure that consists of a collection of inter-connected computing nodes, virtualized computers, and software services that are dynamically provisioned among the competing end-user's applications based on their availability, performance, capability, and Quality of Service (QoS) requirements. Various enterprise Clouds can be pooled together to form a federated infrastructure of resource pools (nodes, services, virtual computers). In a federated organisation: (i) every participant gets access to much larger pools of resources; (ii) the peak-load handling capacity of every enterprise Cloud increases without having the need to maintain or administer any additional computing nodes, services, and storage devices; and (iii) the reliability of a enterprise Cloud is enhanced as a result of multiple redundant clouds that can efficiently tackle disaster condition and ensure business continuity.

Emerging enterprise Cloud applications and the underlying federated hardware infrastructure (Data Centers) are inherently large, with heterogeneous resource types that may exhibit temporal resource conditions. The unique challenges in efficiently managing a federated Cloud computing environment include:

- Large scale – composed of distributed components (services, nodes, applications, users, virtualized computers) that combine together to form a massive environment. These days enterprise Clouds consisting of hundreds of thousands of computing nodes are common (Amazon EC2 [25], Google App Engine [26], Microsoft Live Mesh [27]) and hence federating them together leads to a massive scale environment;

[1] 3rd generation enterprise Grids are exhibiting properties that are commonly envisaged in Cloud computing systems.

- Resource contention - driven by the resource demand pattern and a lack of cooperation among end-user's applications, particular set of resources can get swamped with excessive workload, which significantly undermines the overall utility delivered by the system; and
- Dynamic – the components can leave and join the system at will.

The aforementioned characteristics of the infrastructure accounts to significant development, system integration, configuration, and resource management challenges. Further, the end-users follow a variety of programming models to compose their applications. In other words, in order to efficiently harness the computing power of enterprise Cloud infrastructures [6][25][26][27], software services that can support high level of scalability, robustness, self-organization, and application composition flexibility are required.

This chapter has two objectives. The first is to investigate the challenges as regards to design and development of decentralized, scalable, self-organizing, and federated Cloud computing system. The second is to introduce the Aneka-Federation software system that includes various software services, peer-to-peer resource discovery protocols [5], and resource provisioning methods [3][31] to deal with the challenges in designing decentralized resource management system in a complex, dynamic, and heterogeneous enterprise Cloud computing environment. The components of the Aneka-Federation including computing nodes, services, providers and end-users self-organize themselves based on a structured peer-to-peer routing methodology to create a scalable wide-area overlay of enterprise Clouds. In rest of this chapter, the terms Aneka Cloud(s) and Aneka Enterprise Cloud(s) are used interchangeably.

The unique features of Aneka-Federation are: (i) wide-area scalable overlay of distributed Aneka Enterprise Clouds [6]; (ii) realization of a peer-to-peer based decentralized resource discovery technique as a software service, which has the capability to handle complex resource queries; and (iii) the ability to enforce coordinated interaction among end-users through the implementation of a novel decentralized resource provisioning method. This provisioning method is engineered over a peer-to-peer routing and indexing system that has the ability to route, search and manage complex coordination objects in the system.

The rest of this chapter is organized as follows: Section 2 outlines the challenges and requirements of designing decentralized enterprise Cloud overlays. Section 3 briefly introduces the Aneka Enterprise Cloud system including the basic architecture, key services and programming models. In Section 4, we describe how the Aneka-Federation software system builds upon the decentralized Content-based services. Section 5 lists the comprehensive details on the design and implementation of decentralized Content-based services for message routing, search, and coordinated interaction. Next, Section 6 presents the experimental case study and analysis based on the test run of two enterprise Cloud applications on the Aneka-Federation system. Section 7 puts this work in context with the related works. And finally Section 8 presents a conclusion.

## 2. Designing Decentralized Enterprise Cloud Overlay: Challenges

In decentralized organization of Cloud computing systems both control and decision making are decentralized by nature and where different system components interact together to adaptively maintain and achieve a desired system wide behavior. A distributed Cloud system configuration is considered to be decentralized "*if none of the components in the system are more important than the others, in case that one of the component fails, then it is neither more nor less harmful to the system than caused by the failure of any other component in the system*".

A fundamental challenge in managing the decentralized Cloud computing system is to maintain a consistent connectivity between the components (self-organization) [28]. This challenge cannot be overtaken by introducing a central network model to connect the components, since the information needed for managing the connectivity and making the decisions is completely decentralized and distributed. Further, centralized network model [2] does not scale well, lacks fault-tolerance, and requires expensive server hardware infrastructure.

System components can leave, join, and fail in a dynamic fashion; hence it is an impossible task to manage such a network centrally. Therefore, an efficient decentralized solution is mandatory that can gracefully adapt, and scale to the changing conditions.

A possible way to efficiently interconnect the distributed system components can be based on a structured peer-to-peer overlays. In literature, structured peer-to-peer overlays are more commonly referred to as the Distributed Hash Tables (DHTs). DHTs provide hash table like functionality at the Internet scale. DHTs such as Chord [7], CAN [8], Pastry [15], and Tapestry [9] are inherently self-organizing, fault-tolerant, and scalable. DHTs provide services that are light-weight and hence, do not require an expensive hardware platform for hosting, which is an important requirement as regards to building and managing enterprise Cloud system that consists of commodity machines. A DHT is a distributed data structure that associates a key with a data. Entries in a DHT are stored as a (key, data) pair. A data can be looked up within a logarithmic overlay routing hops if the corresponding key is known.

The effectiveness of the decentralized Cloud computing system depends on the level of coordination and cooperation among the components (users, providers, services) as regards to scheduling and resource allocation. Realizing cooperation among distributed Cloud components requires design and development of the self-organizing, robust, and scalable coordination protocols. The Aneka-Federation system implements one such coordination protocol using the DHT-based routing, lookup and discovery services. The finer details about the coordination protocol are discussed in Section 5.

## 3. Aneka Enterprise Cloud: An Overview

Aneka [6] is a .NET-based service-oriented platform for constructing enterprise Clouds. It is designed to support multiple application models, persistence and security solutions, and communication protocols such that the preferred selection can be changed at anytime without affecting an existing Aneka ecosystem. To create an enterprise Cloud, the resource provider only needs to start an instance of the configurable Aneka container hosting required services on each selected Cloud node. The purpose of the Aneka container is to initialize services and acts as a single point for interaction with the rest of the enterprise Cloud.

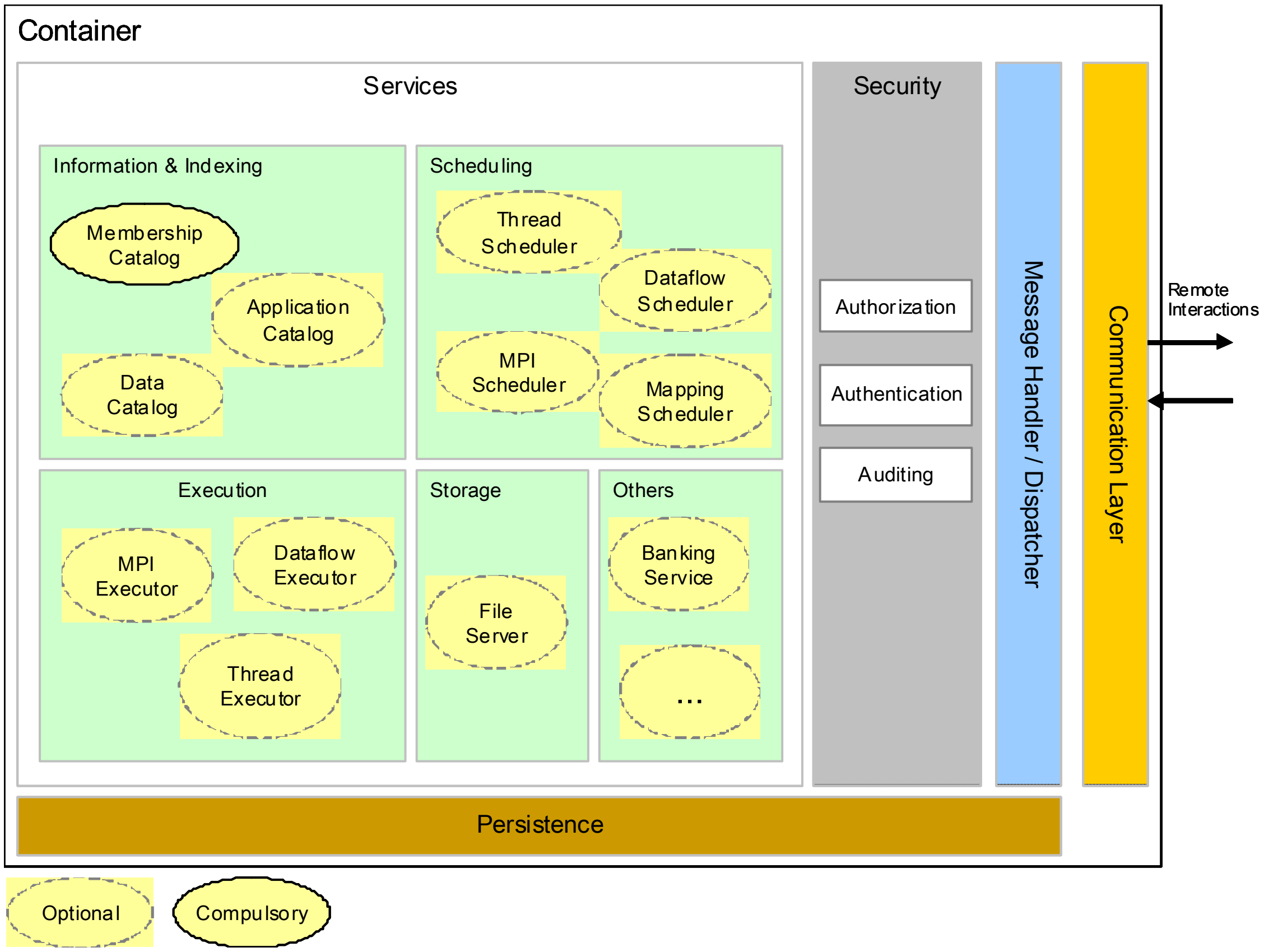

Figure 1: Design of Aneka container.

Figure 1 shows the design of the Aneka container on a single Cloud node. To support scalability, the Aneka container is designed to be lightweight by providing the bare minimum functionality needed for an enterprise Cloud node. It provides the base infrastructure that consists of services for persistence, security (authorization, authentication and auditing), and communication (message handling and dispatching). Every communication within the Aneka services is treated as a message, handled and dispatched through the message handler/dispatcher that acts as a front controller. The Aneka container hosts a compulsory MembershipCatalogue service, which maintains the resource discovery indices (such as a .Net remoting address) of those services currently active in the system. The Aneka container can host any number of optional services that can be added to augment the capabilities of an enterprise Cloud node. Examples of optional services are indexing, scheduling, execution, and storage services. This provides a single, flexible and extensible framework for orchestrating different kinds of enterprise Cloud application models.

To support reliability and flexibility, services are designed to be independent of each other in a container. A service can only interact with other services on the local node or other Cloud node through known interfaces. This means that a malfunctioning service will not affect other working services and/or the container. Therefore, the resource provider can seamlessly configure and manage existing services or introduce new ones into a container. Aneka thus provides the flexibility for the resource provider to implement any network architecture for an enterprise Cloud. The implemented network architecture depends on the interaction of services among enterprise Cloud nodes since each Aneka container on a node can directly interact with other Aneka containers reachable on the network.

## 4. Aneka-Federation

The Aneka-Federation system self-organizes the components (nodes, services, clouds) based on a DHT overlay. Each enterprise Cloud site in the Aneka-Federation (see Figure 2) instantiates a new software service, called Aneka Coordinator. Based on the scalability requirements and system size, an enterprise Cloud can instantiate multiple Aneka Coordinator services. The Aneka Coordinator basically implements the resource management functionalities and resource discovery protocol specifications. The software design of the Aneka-Federation system decouples the fundamental decentralized interaction of participants from the resource allocation policies and the details of managing a specific Aneka Cloud Service. Aneka-Federation software system utilizes the decentralized Cloud services (see Section 5) as regards to efficient distributed resource discovery and coordinated scheduling.

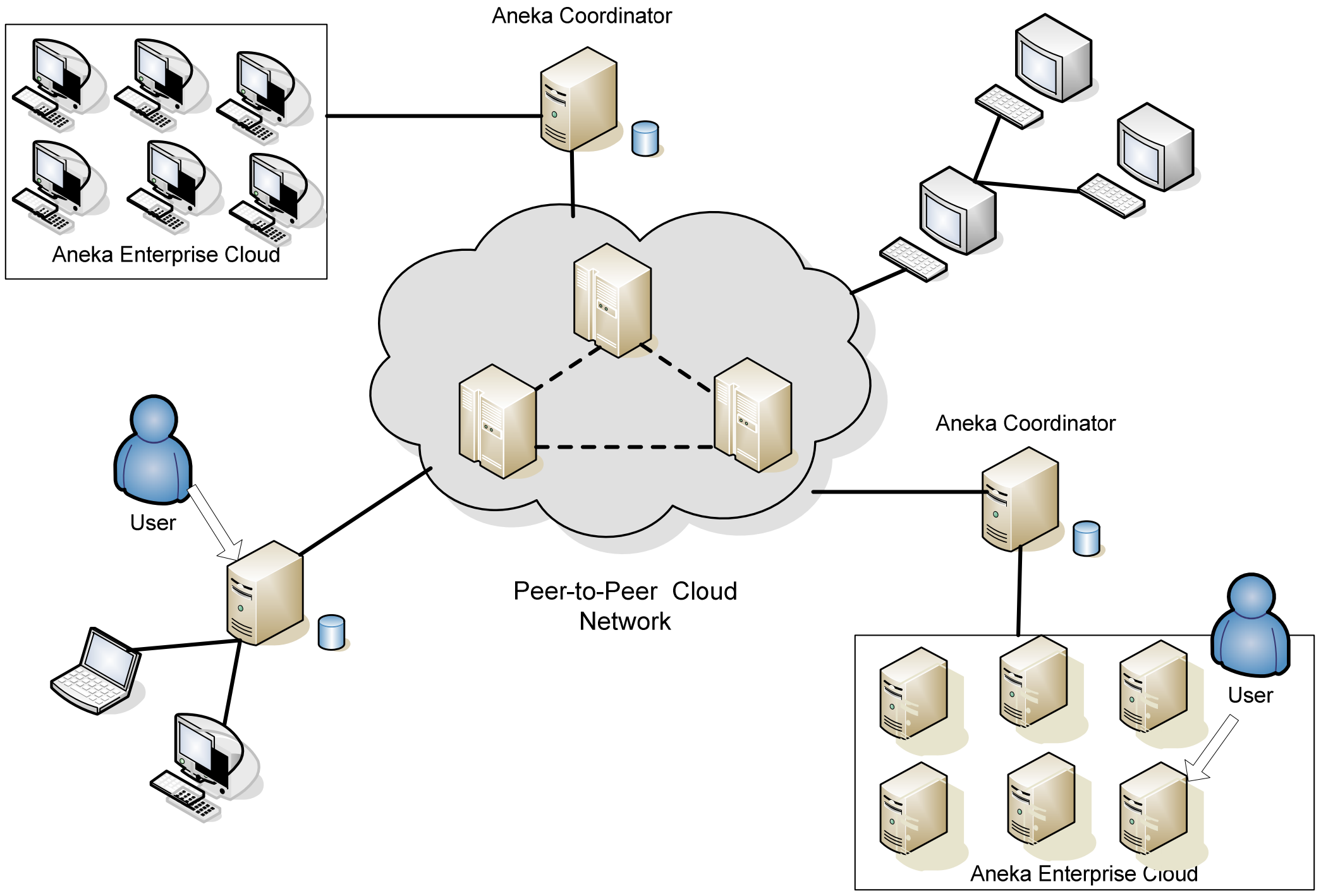

Figure 2: Aneka-Federation network with the Coordinator Services and Aneka Enterprise Clouds.

### 4.1 Design and Implementation

Aneka Coordinator software service is composed of the following components:

- Aneka Services: These include the core services for peer-to-peer scheduling (Thread Scheduler, Task Scheduler, Dataflow Scheduler) and peer-to-peer execution (Thread Executor, Task Executor) provided by the Aneka framework. These services work independently in the container and have the ability to interact with other services such as the P2PMembershipCatalogue through the MessageDispatcher service deployed within each container.
- Aneka Peer: This component of the Aneka Coordinator service loosely glues together the core Aneka services with the decentralized Cloud services. Aneka peer seamlessly encapsulates together the following: Apache Tomcat container (hosting environment and web service front end to the Content-based services), Internet Information Server (IIS) ( hosting environment for ASP.Net service), P2PMembershipCatalogue, and Content-based services (see Figure 4). The basic functionalities of the Aneka peer (refer to Figure 3) include providing services for: (i) Content-based routing of lookup and update messages; and (ii) facilitating decentralized coordination for efficient resource sharing and load-balancing among the Internet-wide distributed Aneka Enterprise Clouds. The Aneka peer service operates at the Core services layer in the layered architecture shown in Figure 9.

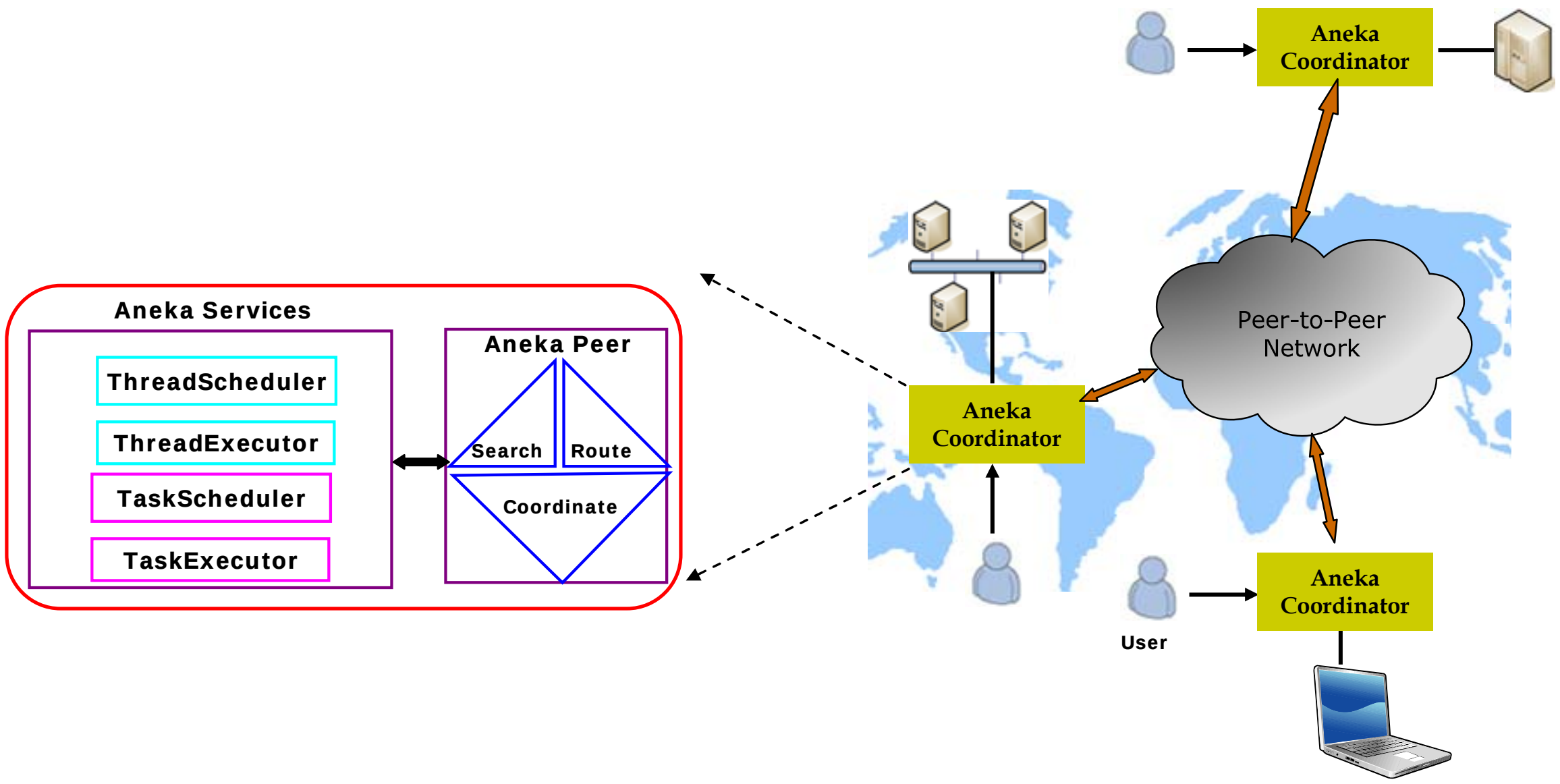

Figure 3: Aneka-Federation over decentralized Cloud services.

Figure 4 shows a block diagram of interaction between various components of Aneka Coordinator software stack. The Aneka Coordinator software stack encapsulates the P2PMembershipCatalogue and Content-based decentralized lookup services. The design components for peer-to-peer scheduling, execution, and membership are derived from the basic Aneka framework components through object oriented software inheritance (see Figures 5, 6, and 7).

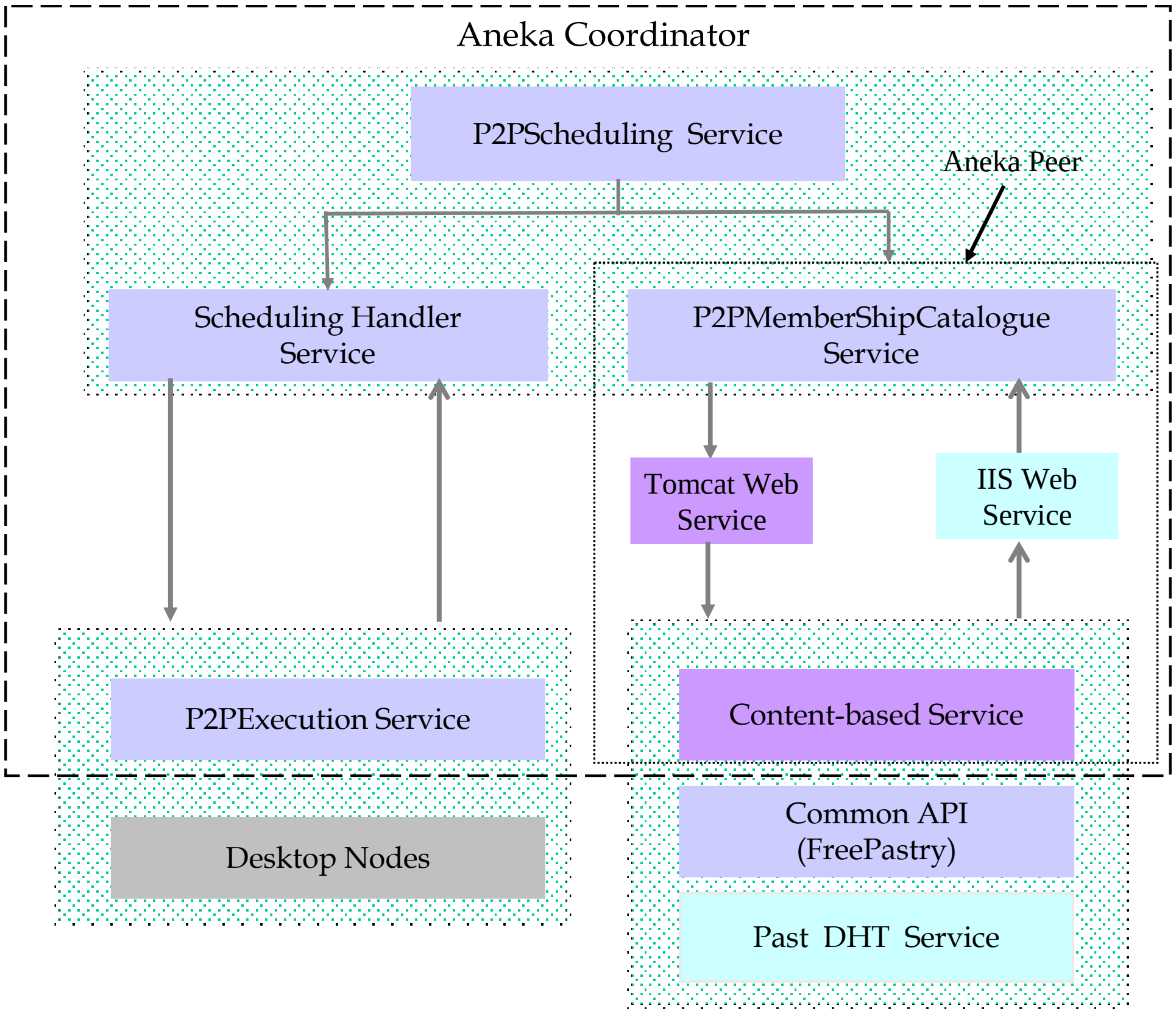

Figure 4: A block diagram showing interaction between various components in the Aneka Coordinator software stack.

A UML (Unified Modeling Language) class diagram that displays the core entities within the Aneka Coordinator's Scheduling service is shown in Figure 5. The main class (refer to Figure 5) that undertakes activities related to application scheduling within the Aneka Coordinator is the P2PScheduling service, which is programmatically inherited from the Aneka's IndependentScheduling service class. The P2PScheduling service implements the methods for: (i) accepting application submission from client nodes (see Figure 8); (ii) sending search query to the P2PMembershipCatalogue service; (iii) dispatching application to Aneka nodes (P2PExecution service); and (iv) collecting the application output data. The core programming models in Aneka including Task, Thread, and Dataflow instantiate P2PScheduling service as their main scheduler class. This runtime binding of P2PScheduling service class to different programming models is taken care of by Microsoft .NET platform and Inverse of Control (IoC) [19] implementation in the Spring .NET framework [20].

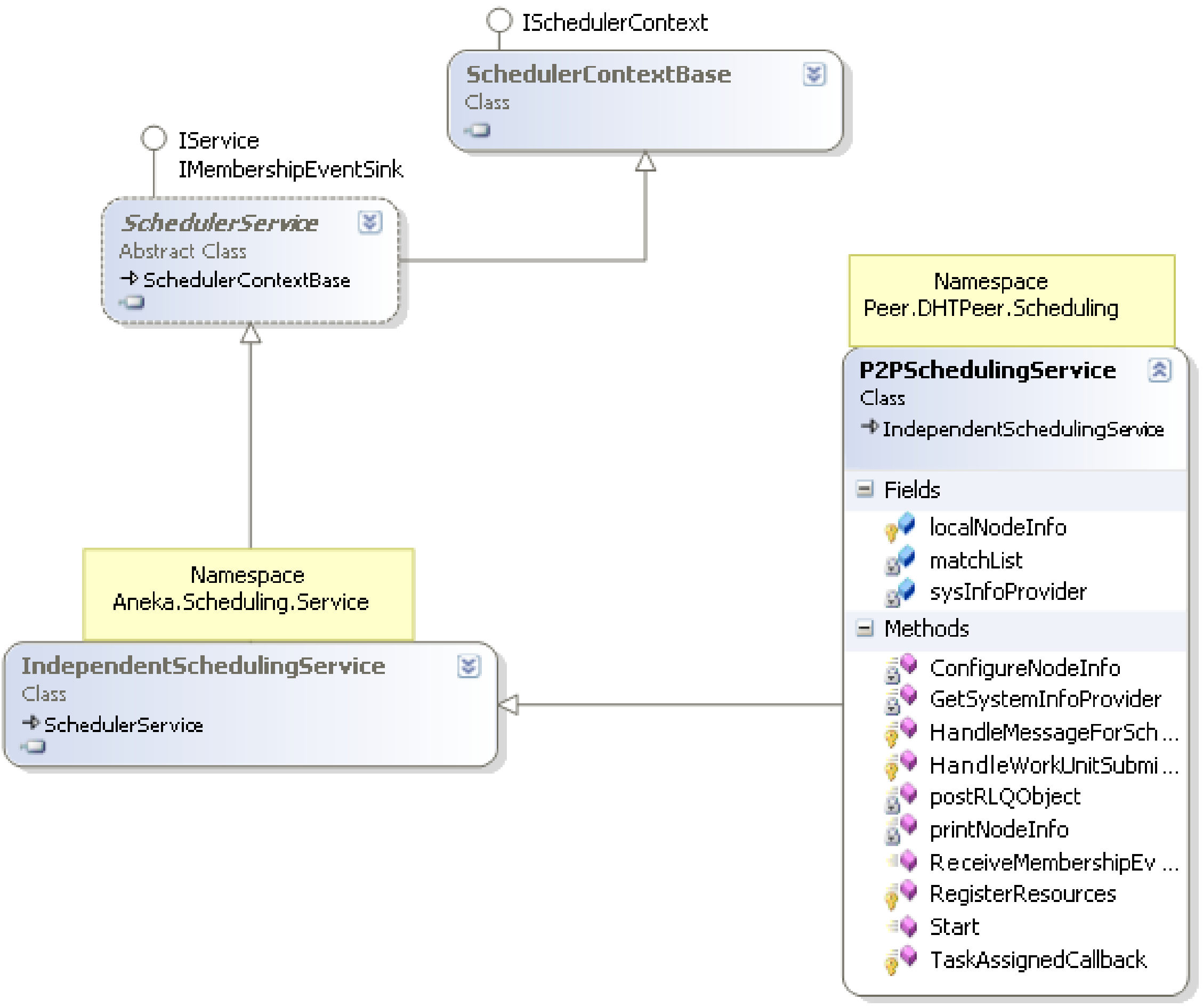

Figure 5: Class design diagram of P2PScheduling service.

Similar to P2PScheduling service, the binding of P2PExecution service to specific programming models (such as P2PTaskExecution, P2PThreadExecution) is done by Microsoft .NET platform and IoC implementation in the Spring .NET framework. The interaction between the services (such as P2PTaskExecution and P2PTaskScheduling service) is facilitated by the MessageDispatcher service. The P2PExecution services update their node usage status with the P2PMembershipCatalogue through the P2PExecutorStatusUpdate component (see Figure 6). The core Aneka Framework defines distinct message types to enable seamless interaction between services. The functionality of handling, compiling, and delivering the messages within the Aneka framework is implemented in the MessageDispactcher service. Recall that the MessageDispatcher service is automatically deployed in the Aneka container.

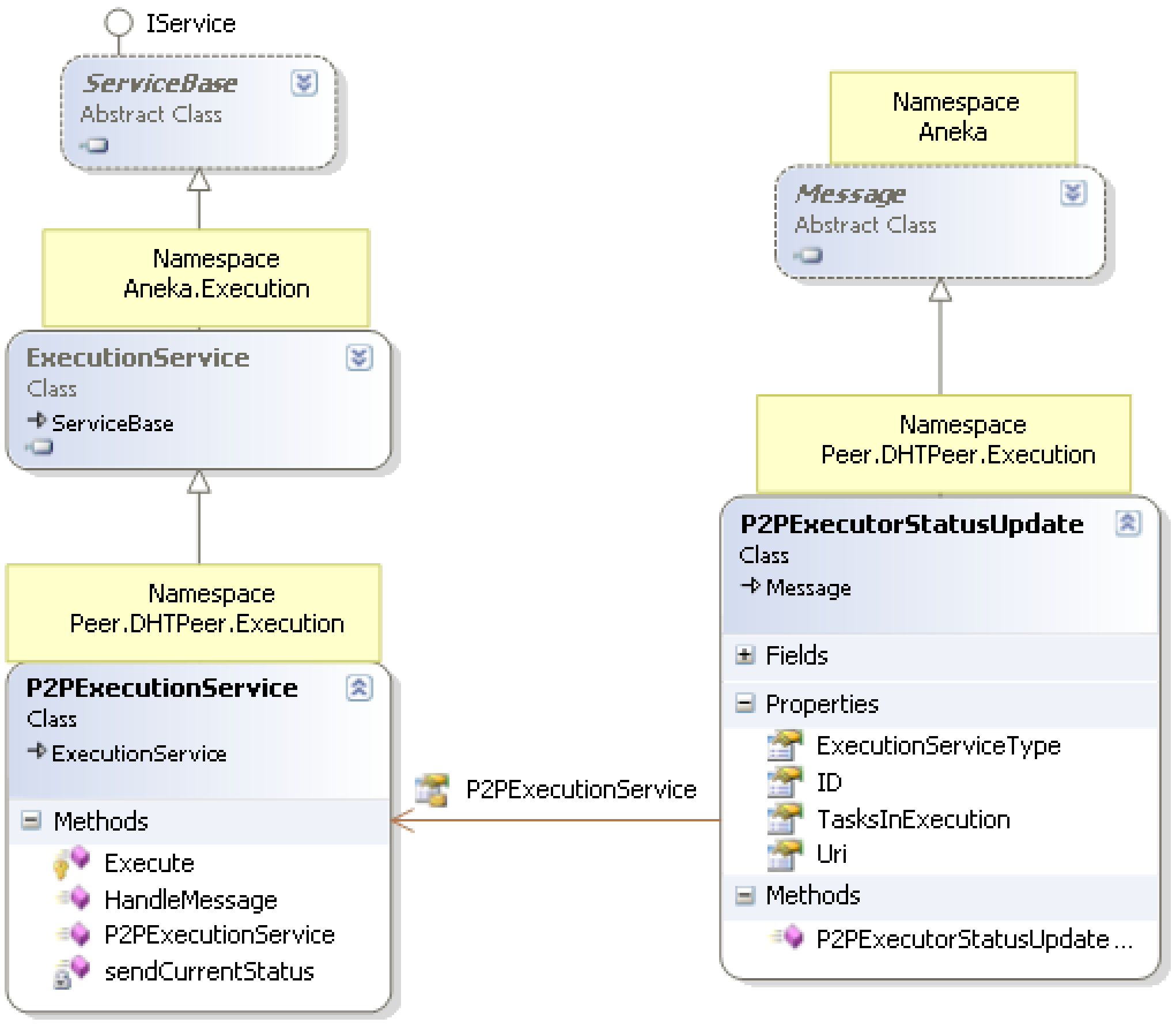

Figure 6: Class design diagram of P2PExecution service.

P2PMembershipCatalogue service is the core component that interacts with the Content-based decentralized Cloud services and aids in the organization and management of Aneka-Federation overlay. The UML class design for this service within the Aneka Coordinator is shown in Figure 7. This service accepts resource claim and ticket objects from P2PScheduling and P2PExecution services respectively (refer to Figure 8), which are then posted with the Content-based services hosted in the Apache Tomcat container.

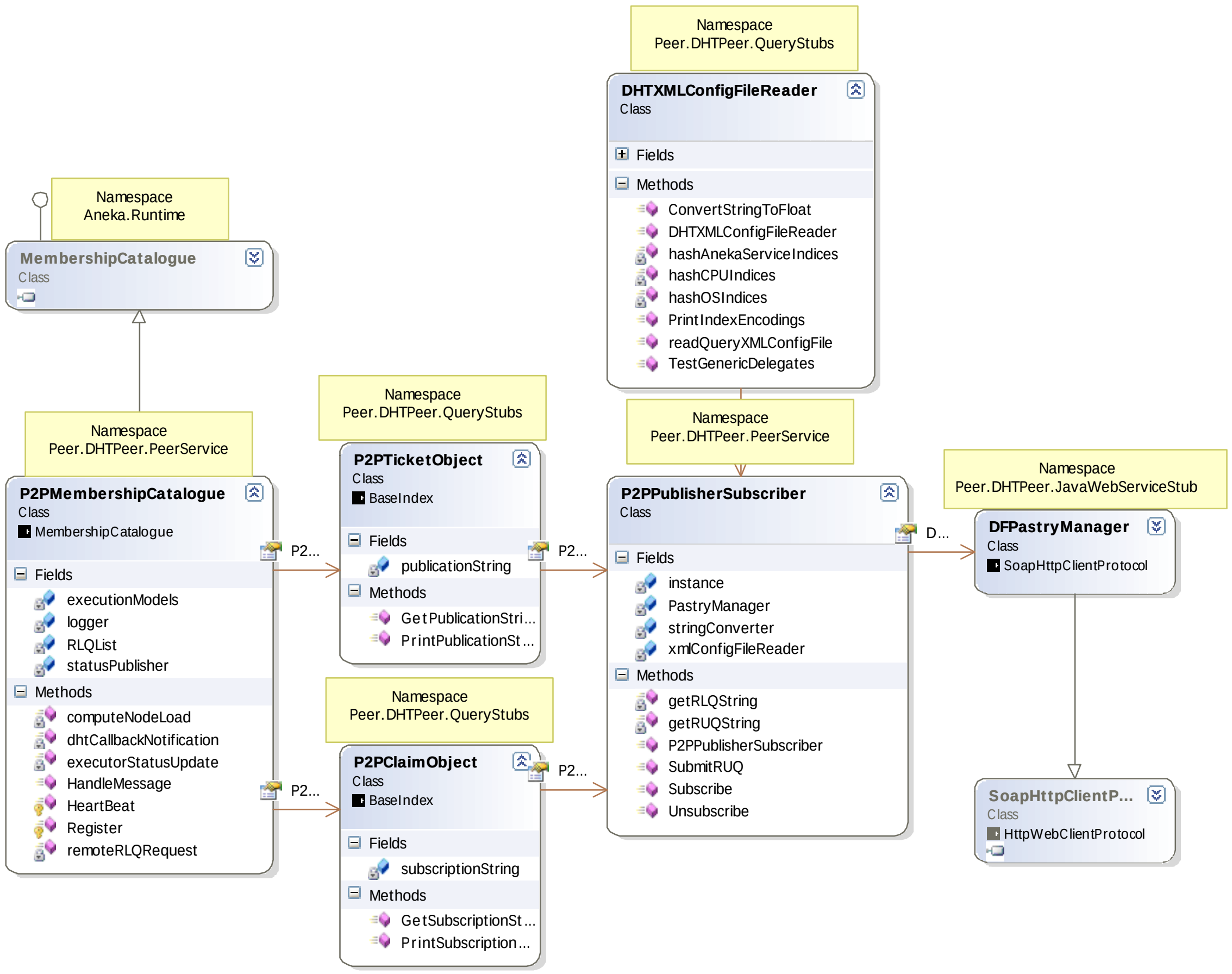

Figure 7: Class design diagram of P2PMembershipCatalogue service.

The P2PMembershipCatalogue interacts with the components hosted within the Apache Tomcat container (Java implementation) using the SOAP-based web services Application Programming Interfaces (APIs) exposed by the DFPastryManager component (see Figure 7). The Content-based service communicates with the P2PMembershipCatalogue service through an ASP.NET web service hosted within in the IIS container (see Figure 4 or 8).

The mandatory services within a Aneka Coordinator that are required to instantiate a fully functional Aneka Enterprise Cloud site includes P2PMembershipCatalogue, P2PExecution, P2PScheduling, .Net web service, and Content-based services (see Figure 8). These services exports a enterprise Cloud site to the federation, and give it capability to accept remote jobs based on its load condition (using their P2PExecution services), and submit local jobs to the federation (through their P2PScheduling services).

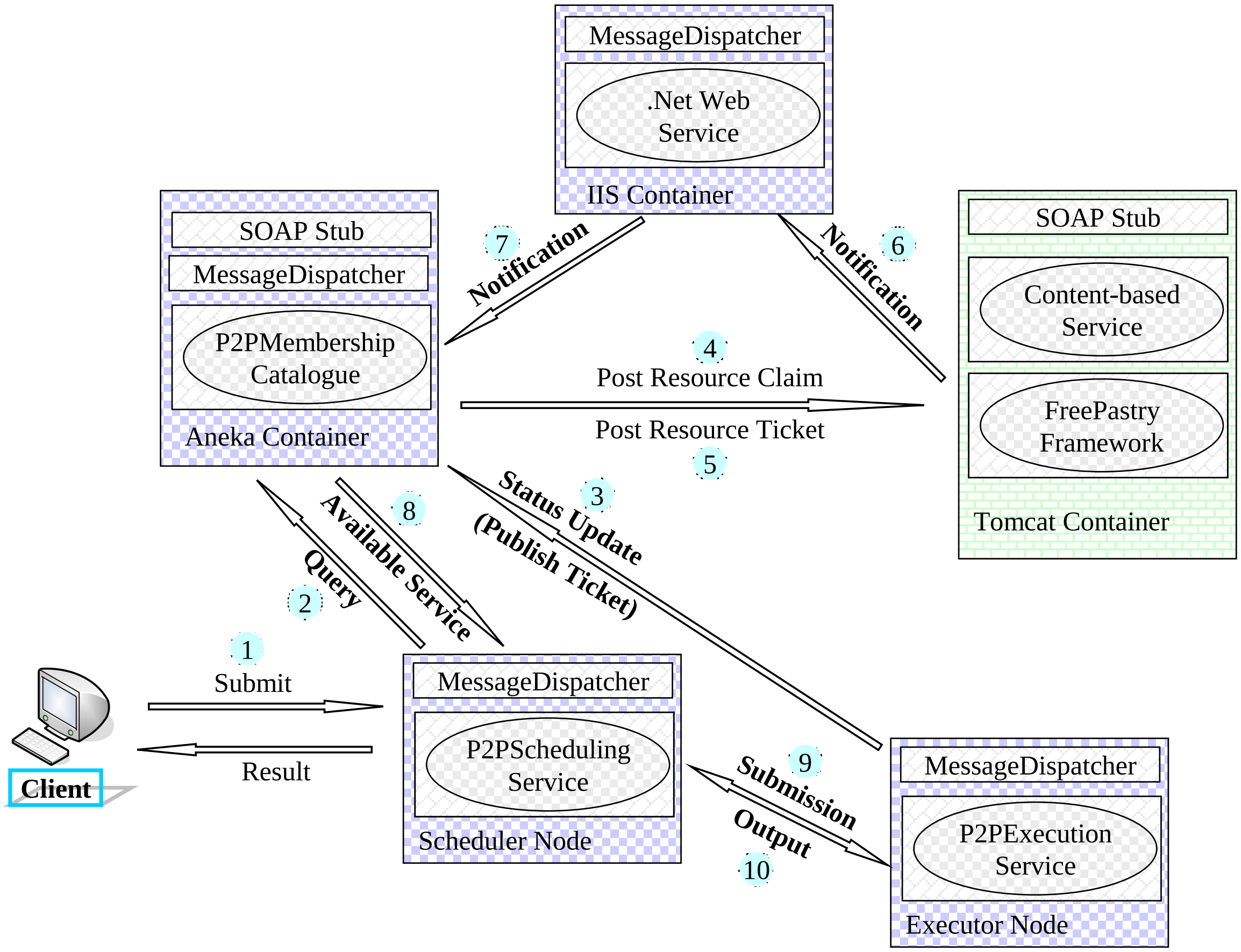

Figure 8: Application execution sequence in Aneka-Federation.

Figure 8 demonstrates a sample application execution flow in the Aneka-Federation system. Clients directly connect and submit their application to a programming model specific scheduling service. For instance, a client having an application programmed using Aneka's

Thread model would submit his application to Thread P2PScheduling service (refer to step 1 in Figure 8). Clients discover the point of contact for local scheduling services by querying their domain specific Aneka Coordinator service. On receipt of an application submission message, a P2PScheduling service encapsulates the resource requirement for that application in a resource claim object and sends a query message to the P2PMembershipCatalogue (see step 2 in Figure 8).

Execution services (such as the P2PThreadExecution, P2PTaskExecution), which are distributed over different enterprise Clouds and administered by enterprise specific Aneka Coordinator services, update their status by sending a resource ticket object to the P2PMembership Catalogue (see step 3 in Figure 8). A resource ticket object in the Aneka-Federation system abstracts the type of service being offered, the underlying hardware platform, and level of QoS that can be supported. The finer details about the composition and the mapping of resource ticket and claim objects are discussed in Section 5.

The P2PMembershipCatalogue then posts the resource ticket and claim objects with the decentralized Content-based services (see step 4 and 5 in Figure 8). When a resource ticket, issued by a P2PTExecution service, matches with a resource claim object, posted by a P2PScheduling service, the Content-based service sends a match notification to the P2PScheduling service through the P2PMembershipCatalogue (see step 6, 7, 8 in figure 8). After receiving the notification, the P2PScheduling service deploys its application on the P2PExecution service (see step 9 in Figure 8). On completion of a submitted application, the P2PExecution service directly returns the output to the P2PScheduling service (see step 10 in Figure 8).

The Aneka Coordinator service supports the following two inter-connection models as regards to an Aneka Enterprise Cloud site creation. First, a resource sharing domain or enterprise Cloud can instantiate a single Aneka-Coordinator service, and let other nodes in the Cloud connect to the Coordinator service. In such a scenario, other nodes need to instantiate only the P2PExecution and P2PScheduling services. These services are dependent on the domain specific Aneka Coordinator service as regards to load update, resource lookup, and membership to the federation (see Figure 11). In second configuration, each node in a resource domain can be installed with all the services within the Aneka Coordinator (see Figure 4). This kind of inter-connection will lead to a true peer-to-peer Aneka-Federation Cloud network, where each node is an autonomous computing node and has the ability to implement its own resource management and scheduling decisions. Hence, in this case the Aneka Coordinator service can support completely decentralized Cloud computing environment both within and between enterprise Clouds.

## 5. Content-based Decentralized Cloud Services

It was pointed out in Section 2 that the DHT based overlay presents a compelling solution for creating a decentralized network of Internet-wide distributed Aneka Enterprise Clouds. However, DHTs are efficient at handling single-dimensional search queries such as "*find all services that match a given attribute value*". Since Cloud computing resources such as enterprise computers, supercomputers, clusters, storage devices, and databases are identified by more than one attribute, therefore a resource search query for these resources is always multi-dimensional. These resource dimensions or attributes include service type, processor speed, architecture, installed operating system, available memory, and network bandwidth. Recent advances in the domain of decentralized resource discovery have been based on extending the existing DHTs with the capability of multi-dimensional data organization and query routing [5].

Our decentralized Cloud management middleware supports peer-to-peer Content-based resource discovery and coordination services for efficient management of distributed enterprise Clouds. The middleware is designed based on a 3-tier layered architecture: the Application layer, Core Services layer, and Connectivity layer (see Figure 9). Cloud services such as the Aneka Coordinator, resource brokers, and schedulers work at the Application layer and insert objects via the Core services layer. The core functionality including the support for decentralized coordinated interaction, and scalable resource discovery is delivered by the Core Services Layer.

The Core services layer, which is managed by the Aneka peer software service, is composed of two sub-layers (see Figure 9): (i) Coordination Service [31]; and (ii) Resource discovery service . The Coordination service component of Aneka peer accepts the coordination objects such as a *resource claim* and *resource ticket*. A resource claim object is a multi-dimensional range look-up query [29] (spatial range object), which is initiated by Aneka Coordinators in the system in order to locate the available Aneka Enterprise Cloud nodes or services that can host their client 's applications. A resource claim object has the following semantics:

*Aneka Service = "P2PThreadExecution" && CPU Type = "Intel" && OSType = "WinXP" && Processor Cores > "1" && Processors Speed > "1.5 GHz"*

On the other hand, a resource ticket is a multi-dimensional point update query (spatial point object), which is sent by an Aneka Enterprise Cloud to report the local Cloud nodes and the deployed services' availability status. A resource ticket object has the following semantics:

*Aneka Service = "P2PThreadExecution" && CPU Type = "Intel" && OSType = "WinXP" && Processor Cores = "2" && Processors Speed = "3 GHz"*

Further, both of these queries can specify different kinds of constraints on the attribute values. If a query specifies a fixed value for each attribute then it is referred to as a multi-dimensional point query. However, in case the query specifies a range of values for attributes, then it is referred to as a multi-dimensional range query. The claim and ticket objects encapsulate coordination logic, which in this case is the resource provisioning logic. The calls between the Coordination service and the Resource Discovery service are made through the standard publish/subscribe technique. Resource Discovery service is responsible for efficiently mapping these complex objects to the DHT overlay.

The Resource Discovery service organizes the resource attributes by embedding a logical publish/subscribe index over a network of distributed Aneka peers. Specifically, the Aneka peer in the system create a DHT overlay that collectively maintain the logical index to facilitate a decentralized resource discovery process. The spatial publish/subscribe index builds a multi-dimensional attribute space based on the Aneka Enterprise Cloud node's resource attributes, where each attribute represents a single dimension. The multi-dimensional spatial index assigns regions of space to the Aneka peer. The calls between Core services layer and Connectivity layer are made through standard DHT primitives such as put(key, value), get(key) that are defined by the peer-to-peer Common Application Programming Interface (API) specification [21].

There are different kinds of spatial indices [5] such as the Space Filling Curves (SFCs) (including the Hilbert curves, Z-curves), k-d tree, MX-CIF Quad tree and R*-tree that can be utilized for managing, routing, and indexing of objects by resource discovery service at Core services layer. Spatial indices are well suited for handling the complexity of Cloud resource queries. Although some spatial indices can have issues as regards to routing load-balance in case of a skewed attribute set, all the spatial indices are generally scalable in terms of the number of hops traversed and messages generated while searching and routing multi-dimensional/spatial claim and ticket objects.

**Resource Claim and Ticket Object Mapping:** At the Core services layer, a spatial index that assigns regions of multi-dimensional attribute space to Aneka peers has been implemented. The MX-CIF Quadtree spatial hashing technique [23] is used to map the logical multi-dimensional control point (point C in Figure 10 represents a *2*-dimensional control point) onto a Pastry DHT overlay. If an Aneka peer is assigned a region in the multi-dimensional attribute space, then it is responsible for handling all the activities related to the lookups and updates that intersect with the region of space. Figure 10 depicts a *2*-dimensional Aneka resource attribute space for mapping resource claim and ticket objects. The attribute space resembles a mesh-like structure due to its recursive division process. The index cells, resulted from this process, remain constant throughout the life of a *d*-dimensional attribute space and serve as the entry points for subsequent mapping of claim and ticket objects. The number of index cells produced at the minimum division level, $f_{min}$ is always equal to $(f_{min})^{dim}$, where *dim* is the dimensionality of the attribute space. These index cells are called *base index cells* and they are initialized when the Aneka Peers bootstrap to the federation network. Finer details on the recursive subdivision technique can be found in [23]. Every Aneka Peer in the federation has the basic information about the attribute space coordinate values, dimensions and minimum division levels.

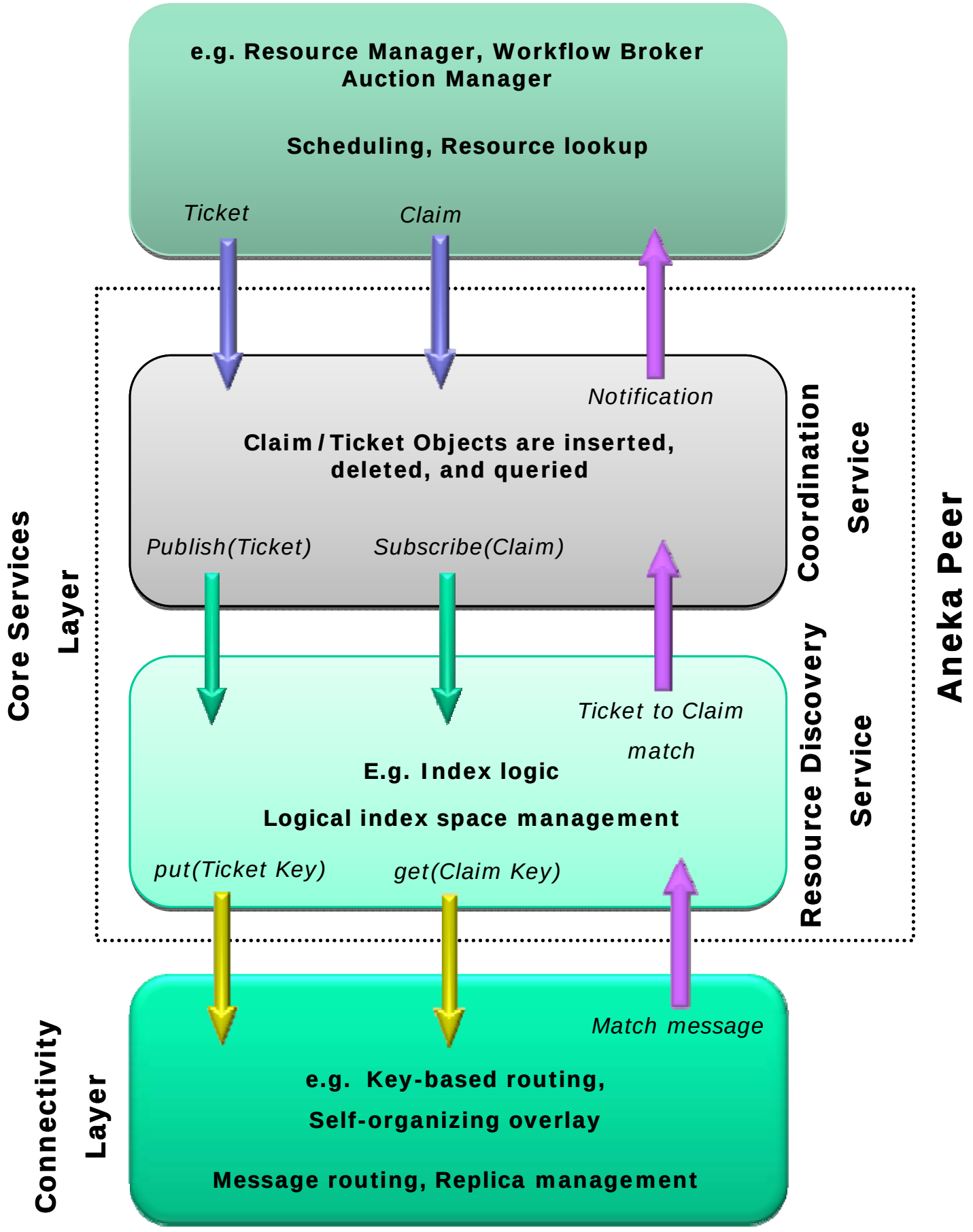

Figure 9: Layered view of the Content-based decentralized Cloud services.

Every cell at the $f_{min}$ level is uniquely identified by its centroid, termed as the *control point*. Figure 10 shows four control points A, B, C, and D. A DHT hashing (cryptographic functions such as SHA-1/2) method is utilized to map the responsibility of managing control points to the Aneka Peers. In a *2*-dimensional setting, an index cell $i$ = ($x_1$, $y_1$, $x_2$, $y_2$), and its control point are computed as (($x_2$-$x_1$)/2, ($y_2$-$y_1$)/2). The spatial hashing technique takes two input parameters, SpatialHash (control point coordinates, object's coordinates), in terms of DHT common API primitive that can be written as Put (Key, Value), where the cryptographic hash of the control point acts as the Key for DHT overlay, while Value is the coordinate values of the resource claim or ticket object to be mapped. In Figure 10, the Aneka peer at Cloud *s* is assigned index cell *i* through the spatial hashing technique, which makes it responsible for managing all objects that map to the cell *i* (Claim T2, T3, T4 and Ticket s).

For mapping claim objects, the process of mapping index cells to the Aneka Peers depends on whether it is spatial point object or spatial range object. The mapping of point object is simple since every point is mapped to only one cell in the attribute space. For spatial range object (such as Claims T2, T3 or T4), the mapping is not always singular because a range object can cross more than once index cell (see Claim T5 in Figure 10). To avoid mapping a spatial range object to all the cells that it intersects, which can create many duplicates, a mapping strategy based on diagonal hyperplane [24] in the attribute space is implemented. This mapping involves feeding spatial range object coordinate values and candidate index as inputs to a mapping function, $F_{map}$ (spatial object, candidate index cells). An Aneka Peer service uses the index cell(s) currently assigned to it and a set of known base index cells as candidate cells, which are obtained at the time of bootstrapping into the federation. The $F_{map}$ returns the index cells and their control points to which the given spatial range object should be stored with. Next, these control points and the spatial object is given as inputs to function SpatialHash(control point, object), which in connection with the Connectivity layer generates DHT Ids (Keys) and performs routing of claim/ticket objects to the Aneka Peers.

Similarly, the mapping process of a ticket object also involves the identification of the intersection index cells in the attribute space. A ticket is always associated with a region [24] and all cells that fall fully or partially within the region are selected to receive the corresponding ticket. The calculation of the region is based upon the diagonal hyperplane of the attribute space.

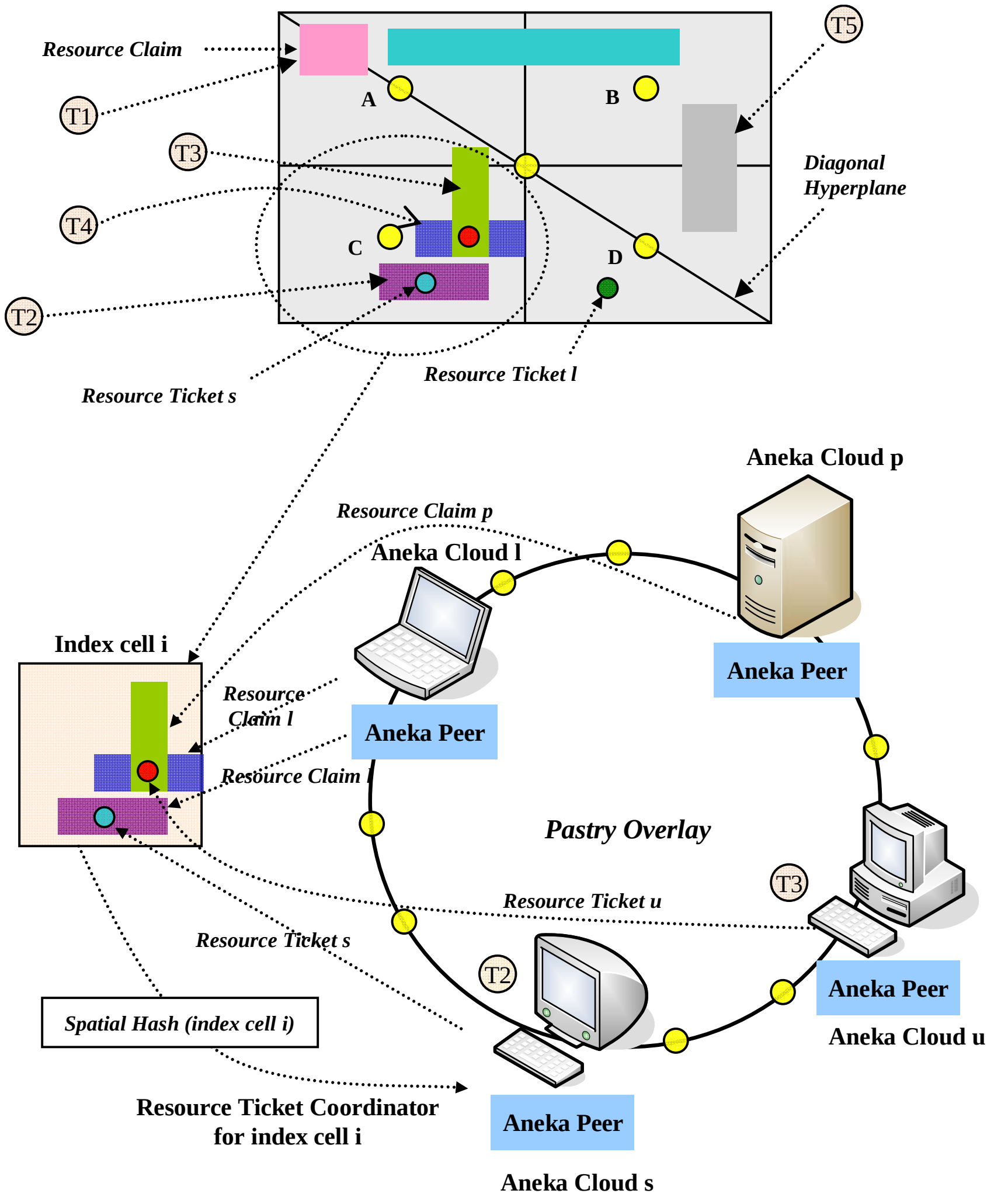

Figure 10: Resource claim and ticket object mapping and coordinated scheduling across Aneka Enterprise Cloud sites. Spatial resource claims {T1, T2, T3, T4}, index cell control points {A, B, C, D}, spatial point tickets {l, s} and some of the spatial hashings to the Pastry ring, i.e. the *d*-dimensional (spatial) coordinate values of a cell's control point is used as the Pastry key. For this Figure $f_{min}$ =2, dim = 2.

**Coordinated Load Balancing**: Both resource claim and ticket objects are spatially hashed to an index cell *i* in the multi-dimensional Aneka services' attribute space. In Figure 10, resource claim object for task T1 is mapped to index cell A, while for T2, T3, and T4, the responsible cell is *i* with control point value C. Note that, these resource claim objects are posted by P2PScheduling services (Task or Thread) of Aneka Cloud nodes. In Figure 10, scheduling service at Cloud *p* posts a resource claim object which is mapped to index cell *i*. The index cell *i* is *spatially* hashed to an Aneka peer at Cloud *s*. In this case, Cloud *s* is responsible for coordinating the resource sharing among all the resource claims that are currently mapped to the cell *i*. Subsequently, Cloud *u* issues a resource ticket (see Figure 10) that falls under a region of the attribute space currently required by the tasks T3 and T4. Next, the coordination service of Aneka peer at Cloud *s* has to decide which of the tasks (either T3 or T4 or both) is allowed to claim the ticket issues by Cloud *u*. The load-balancing decision is based on the principle that it should not lead to over-provisioning of resources at Cloud *u*. This mechanism leads to coordinated load-balancing across Aneka Enterprise Clouds and aids in achieving system-wide objective function, while at the same time preserving the autonomy of the participating Aneka Enterprise Clouds.

Table 1: Claims stored with an Aneka Peer service at time *T*.

| **Time** | **Claim ID** | **Service Type** | **Speed (GHz)** | **Processors** | **Type** |
|---|---|---|---|---|---|
| 300 | Claim 1 | P2PThreadExecution | > 2 | 1 | Intel |
| 400 | Claim 2 | P2PTaskExecution | > 2 | 1 | Intel |
| 500 | Claim 3 | P2PThreadExecution | > 2.4 | 1 | Intel |

Table 2: Ticket Published with an Aneka Peer service at time *T*.

| **Time** | **Cloud ID** | **Service Type** | **Speed (GHz)** | **Processors** | **Type** |
|---|---|---|---|---|---|
| 700 | Cloud 2 | P2PThreadExecution | 2.7 | 1 (available) | Intel |

The examples in Table 1 are list of resource claim objects that are stored with an Aneka peer's coordination service at time *T* = 700 secs. Essentially, the claims in the list arrived at a time <= 700 and wait for a suitable ticket object that can meet their application's requirements (software, hardware, service type). Table 2 depicts a ticket object that has arrived at *T* = 700.

Following the ticket arrival, the coordination service undertakes a procedure that allocates the ticket object among the list of matching claims. Based on the Cloud node's attribute specification, both Claim 1 and Claim 2 match the ticket issuing Cloud node's configuration. As specified in the ticket object, there is currently one processor available within the Cloud 2, which means that at this time only Claim 1 can be served. Following this, the coordination service notifies the Aneka-Coordinator, which has posted the Claim 1. Note that Claims 2 and 3 have to wait for the arrival of tickets that can match their requirements.

The Connectivity layer is responsible for undertaking a key-Based routing in the DHT overlay, where it can implement the routing methods based on DHTs, such as Chord, CAN, and Pastry. The actual implementation protocol at this layer does not directly affect the operations of the Core services layer. In principle, any DHT implementation at this layer could perform the desired task. DHTs are inherently self-organizing, fault-tolerant, and scalable.

At the Connectivity layer, our middleware utilizes the open source implementation of Pastry DHT known as the FreePastry [30]. FreePastry offers a generic, scalable and efficient peer-to-peer routing framework for the development of decentralized Cloud services. FreePastry is an open source implementation of well-known Pastry routing substrate. It exposes a Key-based Routing (KBR) API and given the Key K, Pastry routing algorithm can find the peer responsible for this key in $\log_b n$ messages, where *b* is the base and *n* is the number of Aneka Peers in the network. Nodes in a Pastry overlay form a decentralized, self-organising and fault-tolerant circular network within the Internet. Both data and peers in the Pastry overlay are assigned Ids from 160-bit unique identifier space. These identifiers are generated by hashing the object's names, a peer's IP address or public key using the cryptographic hash functions such as SHA-1/2. FreePastry is currently available under BSD-like license. FreePastry framework supports the P2P Common API specification proposed in the paper [21].

## 6. Experimental Evaluation and Discussion

In this section, we evaluate the performance of the Aneka-Federation software system by creating a resource sharing network that consists of 5 Aneka Enterprise Clouds (refer to Figure 11). These Aneka Enterprise Clouds are installed and configured in three different Laboratories (Labs) within the Computer Science and Software Engineering Department, The University of Melbourne. The nodes in these Labs are connected through a Local Area Network (LAN). The LAN connection has a data transfer bandwidth of 100 Mb/Sec (megabits per seconds). Next, the various parameters and application characteristics related to this study are briefly described.

**Aneka Enterprise Cloud Configuration**: Each Aneka Cloud in the experiments is configured to have 4 nodes out of which, one of the nodes instantiates the Aneka-Coordinator service. In addition to the Aneka Coordinator service, this node also hosts the other optional services including the P2PScheduling (for Thread and Task models) and P2PExecution services (for Thread and Task models). The remaining 3 nodes are configured to run the P2PExecution services for Task and Thread programming models. These nodes connect and communicate with the Aneka-Coordinator service through .Net remoting messaging APIs. The P2PExecution services periodically update their usage status with the Aneka-Coordinator service. The update delay is configurable parameter with values in milliseconds or seconds. The nodes across different Aneka Enterprise Clouds update their status dynamically with the decentralized Content-based services. The node status update delays across the Aneka Enterprise Clouds are uniformly distributed over interval [5, 40] seconds.

**FreePastry Network Configuration**: Both Aneka Peers' nodeIds and claim/ticket objectIds are randomly assigned from and uniformly distributed in the 160-bit Pastry identifier space. Every Content-based service is configured to buffer maximum of 1000 messages at a given instance of time. The buffer size is chosen to be sufficiently large such that the FreePastry does not drop any messages. Other network parameters are configured to the default values as given in the file freepastry.params. This file is provided with the FreePastry distribution.

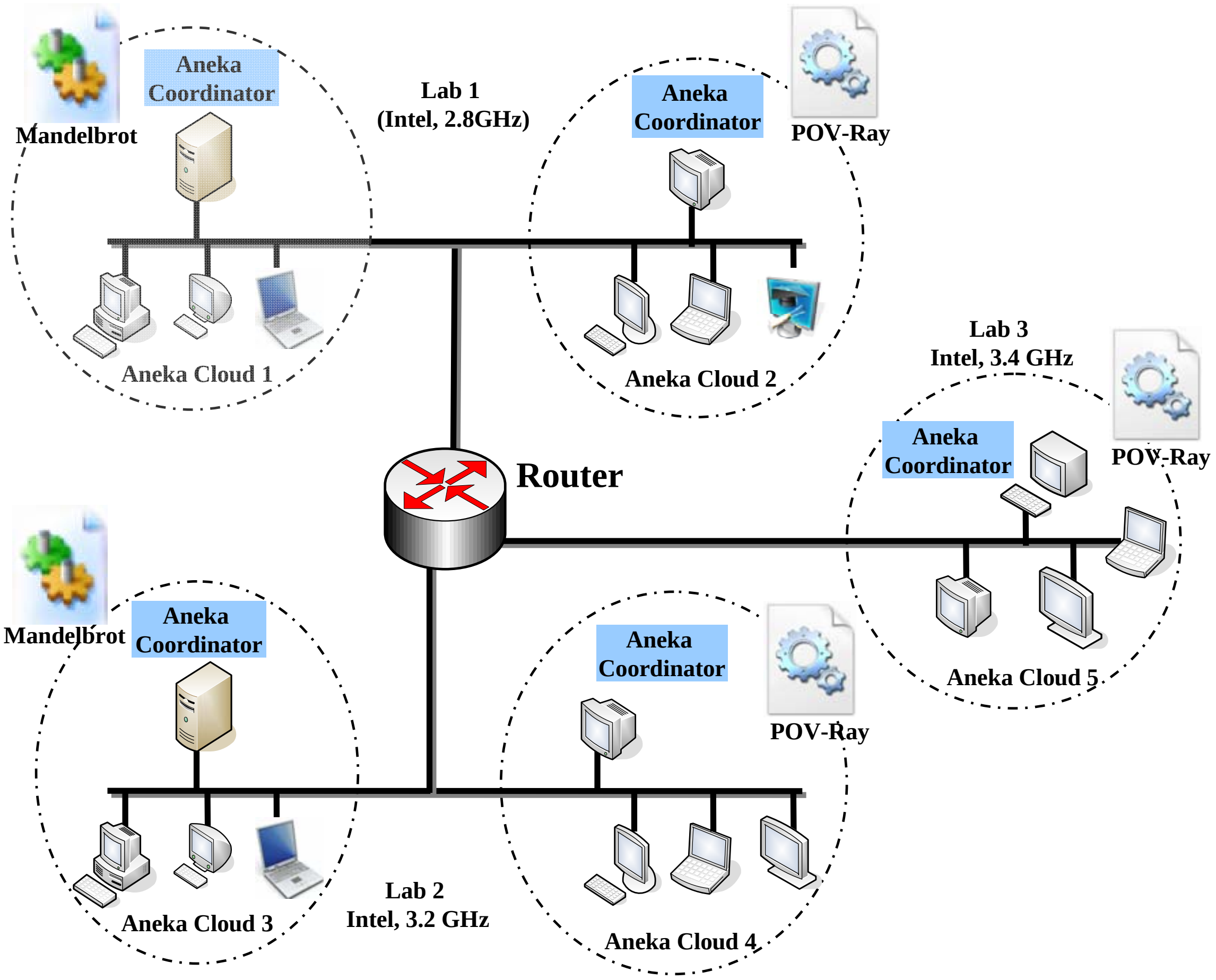

Figure 11: Aneka-Federation test bed distributed over 3 departmental laboratories.

**Spatial Index Configuration**: The minimum division $f_{min}$ of logical *d*-dimensional spatial index that forms the basis for mapping, routing, and searching the claim and ticket objects is set to 3, while the maximum height of the spatial index tree, $f_{max}$ is constrained to 3. In other words, the division of the *d*-dimensional attribute is not allowed beyond $f_{min.}$ This is done for simplicity, understanding the load balancing issues of spatial indices [23] with increasing $f_{max}$ is a different research problem and is beyond scope of this chapter. The index space has provision for defining claim and ticket objects that specify the Aneka nodes/service's characteristics in *4* dimensions including number of Aneka service type, processors, processor architecture, and processing speed. The aforementioned spatial index configuration results into $81(3^4)$ index cells at $f_{min}$ level. On an average, 16 index cells are hashed to an Aneka Peer in a network of 5 Aneka Coordinators.

**Claim and Ticket Object's Spatial Extent**: Ticket objects in the Aneka-Federation express equality constraints on an Aneka node's hardware/software attribute value (e.g. =). In other words, ticket objects are always *d*-dimensional(spatial) point query for this study. On the other hand, the claim objects posted by P2PScheduling services have their spatial extent in *d* dimensions with both, range and fixed constraint (e.g. >=, <=) for the attributes. The spatial extent of a claim object in different attribute dimension is controlled by the characteristic of the node, which is hosting the P2PScheduling service. Attributes including Aneka service type, processor architecture, and number of processors are fixed, i.e. they are expressed as equality constraints. The value for processing speed is expressed using >= constraints, i.e. search for the Aneka services, which can process application atleast as fast as what is available on the submission node. However, the P2PScheduling services can create claim objects with different kind of constraints, which can result in different routing, searching, and matching complexity. Studying this behavior of the system is beyond the scope of this chapter.

**Application Models**: Aneka supports composition and execution of application programmers using different models [22] to be executed on the same enterprise Cloud infrastructure. The experimental evaluation in this chapter considers simultaneous execution of applications programmed using Task and Thread models. The Task model defines an application as a collection of one or more tasks, where each task represents an independent unit of execution. Similarly, the Thread model defines an application as a collection of one or more independent threads. Both models can be successfully utilized to compose and program embarrassingly parallel programs (parameter sweep applications). The Task model is more suitable for cloud enabling the legacy applications, while the Thread model fits better for implementing and architecting new applications, algorithms on clouds since it gives finer degree of control and flexibility as regards to runtime control.

To demonstrate the effectiveness of the Aneka-Federation platform with regards to: (i) ease of heterogeneous application composition flexibility; (ii) different programming model supportability; and (iii) concurrent scheduling feasibility of heterogeneous applications on shared Cloud computing infrastructure, the experiments are run based on the following applications:

- **Persistence of Vision Raytracer (POV-Ray [33])**: This application is cloud enabled using the Aneka Task programming model. POV-Ray is an image rendering application that can create very complex and realistic three dimensional models. Aneka POV-Ray application interface allows the selection of a model, the dimension of the rendered image, and the number of independent tasks into which rendering activities have to be partitioned. The task partition is based on the values that a user specifies for parameter rows and columns on the interface. In the experiments, the values for the rows and the columns are varied over the interval [5 x 5, 13 x 13] in steps of 2.
- **Mandelbrot [34]**: Mathematically, the Mandelbrot set is an ordered collection of points in the complex plane, the boundary of which forms a fractal. Aneka implements and cloud enables the Mandelbrot fractal calculation using the Thread programming model. The application submission interface allows the user to configure number of horizontal and vertical partitions into which the fractal computation can be divided. The number of independent thread units created is equal to the horizontal x vertical partitions. For evaluations, we vary the values for horizontal and vertical parameters over the interval [5 x 5, 13 x 13 ] in steps of 2. This configuration results in 5 observation points.

**Results and Discussion**

To measure the performance of Aneka-Federation system as regards to scheduling, we quantify the response time metric for the POV-Ray and Mandelbrot applications. The response time for an application is computed by subtracting the output arrival time of the last task/thread in the execution list from the time at which the application is submitted. The observations are made for different application granularity(sizes) as discussed in the last Section.

Figure 12 depicts the results for response time in seconds with increasing granularity for the POV-Ray application. The users at Aneka Cloud 1, 3, 4 submit the applications to their respective Aneka Coordinator services (refer to the Figure 11). The experiment results show that the POV-Ray application submitted at Aneka Cloud 1 experienced comparatively lesser response times for its POV-Ray tasks as compared to the ones submitted at Aneka Cloud 3 and 4. The fundamental reasons behind this behavior of system is that the spatial extent and attribute constraints of the resource claim objects posted by the P2PTaskScheduling service at Aneka Cloud 1. As shown in Figure 11, every Aneka Cloud offers processors of type "Intel" with varying speed. Based on the in the previous Section, the processing speed is expressed using >= constraints, which means that the application submitted in the Aneka Enterprise Clouds, 1 and 2 (processing speed = 2.4 GHz), can be executed on any of the nodes in the enterprise Clouds 1, 2, 3, 4, and 5.

However, the application submitted at Aneka Clouds 3 and 4 can be executed only on Clouds 3, 4, and 5. Accordingly, the application submitted in Aneka Cloud 3 can only be processed locally as the spatial dimension and processing speed for the resource claim objects specifies constraints as >= 3.5 GHz. Due to these spatial constraints on the processing speed attribute value, the application in different Clouds gets access to varying Aneka node pools that result in different levels of response times. For the aforementioned arguments, it can be seen in Figure 12 and 13 (Mandelbrot applications) that applications at Aneka Clouds 1 and 2 have relatively better response times as compared to the ones submitted at Aneka Cloud 3, 4, and 5.

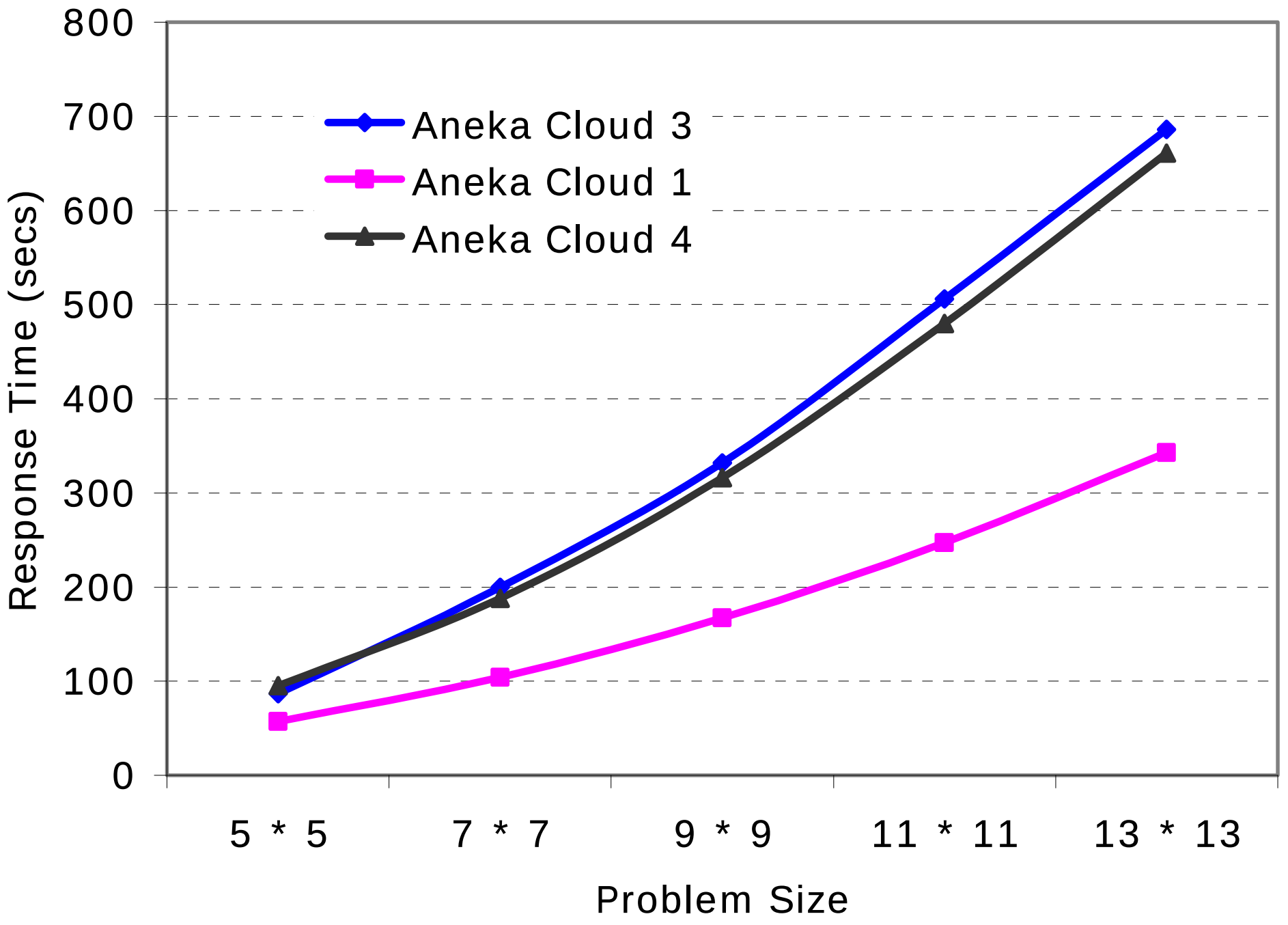

Figure 12: POV-Ray Application: Response Time (secs) vs. Problem Size

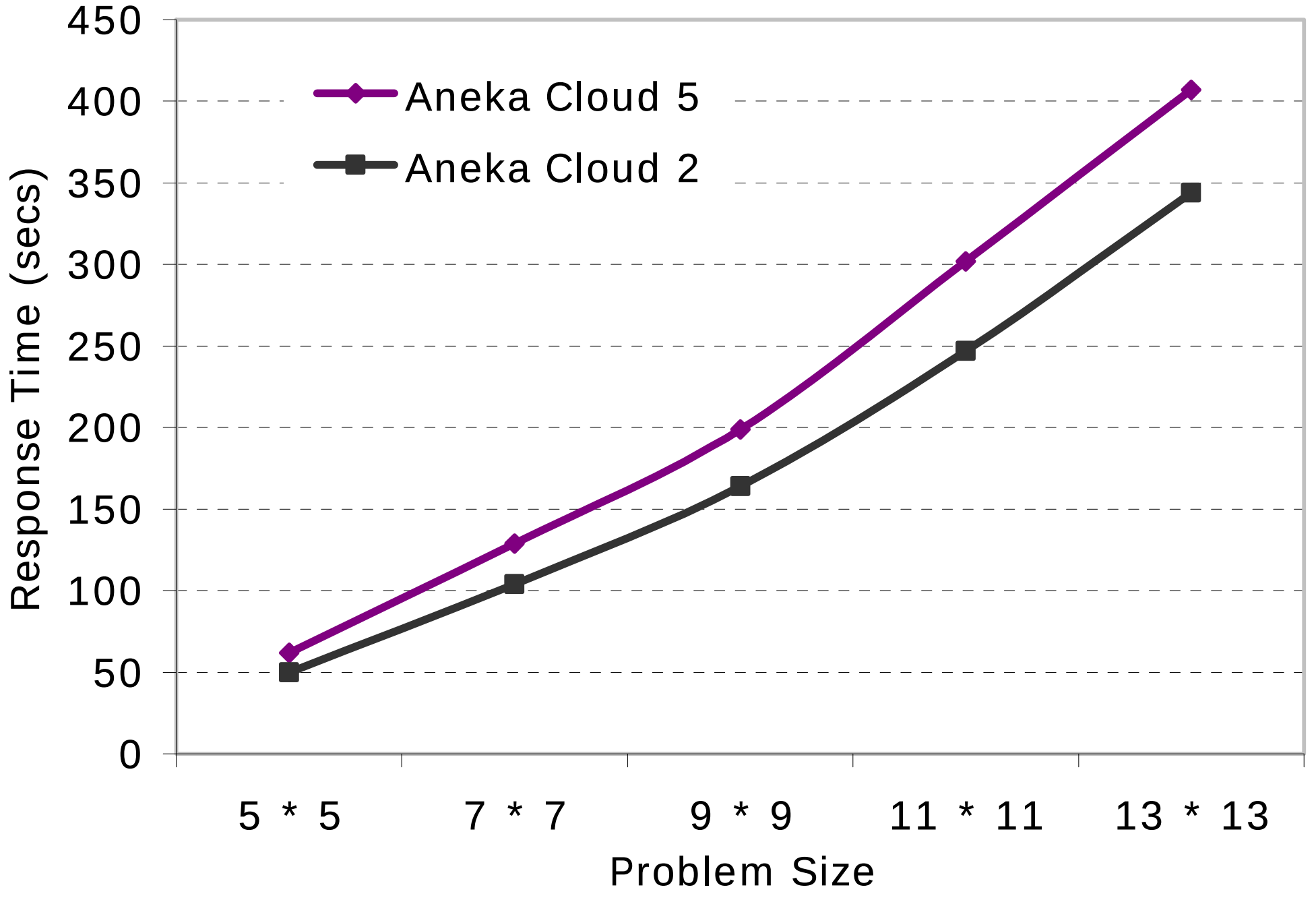

Figure 13: Mandelbrot Application: Response Time (Secs) vs. Problem Size.

Figures 14 and 15 present the results for the total number of jobs processed in different Aneka Clouds by their P2PTaskExecution and P2PThreadExectuion services. The results show that the P2PTaskExecution and P2PThreadExecution services hosted within the Aneka Clouds 3, 4, and 5 processes relatively more jobs as compared to those hosted within Aneka Clouds 1 and 2. This happens due to the spatial constraint on the processing speed attribute value in the resource claim object posted by different P2PScheduling (Task/Thread) services across the Aneka Clouds. As Aneka Cloud 5 offers the fastest processing speed (within the spatial extent of all resource claim objects in the system), it processes more jobs as compared to other Aneka Clouds in the federation (see Figure 14 and 15). Thus, in the proposed Aneka-Federation system, the spatial extent for resource attribute values specified by the P2PScheduling services directly controls the job distribution and application response times in the system.

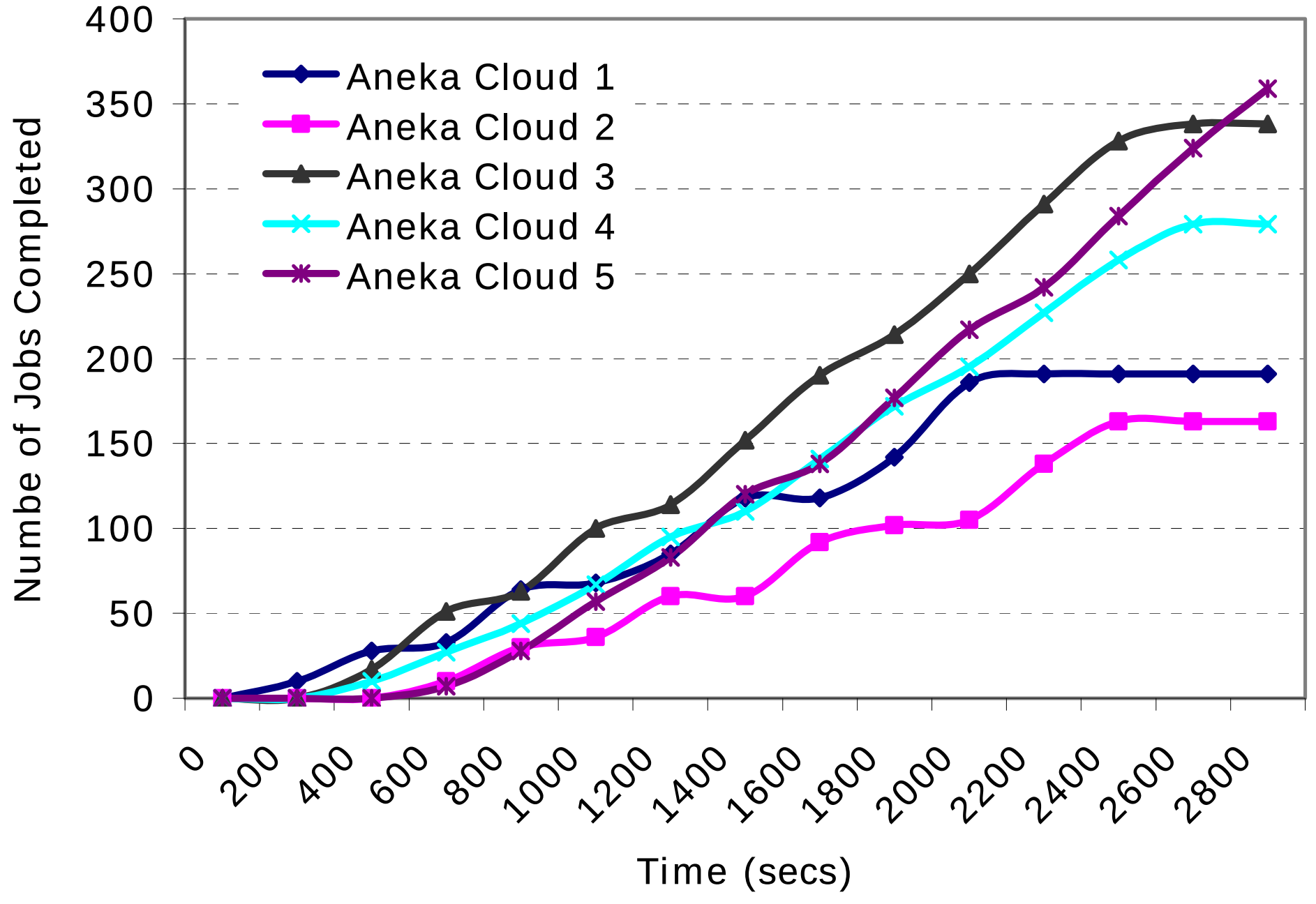

Figure 14: P2PTaskExecution Service: Time (secs) vs. Number of Jobs Completed.

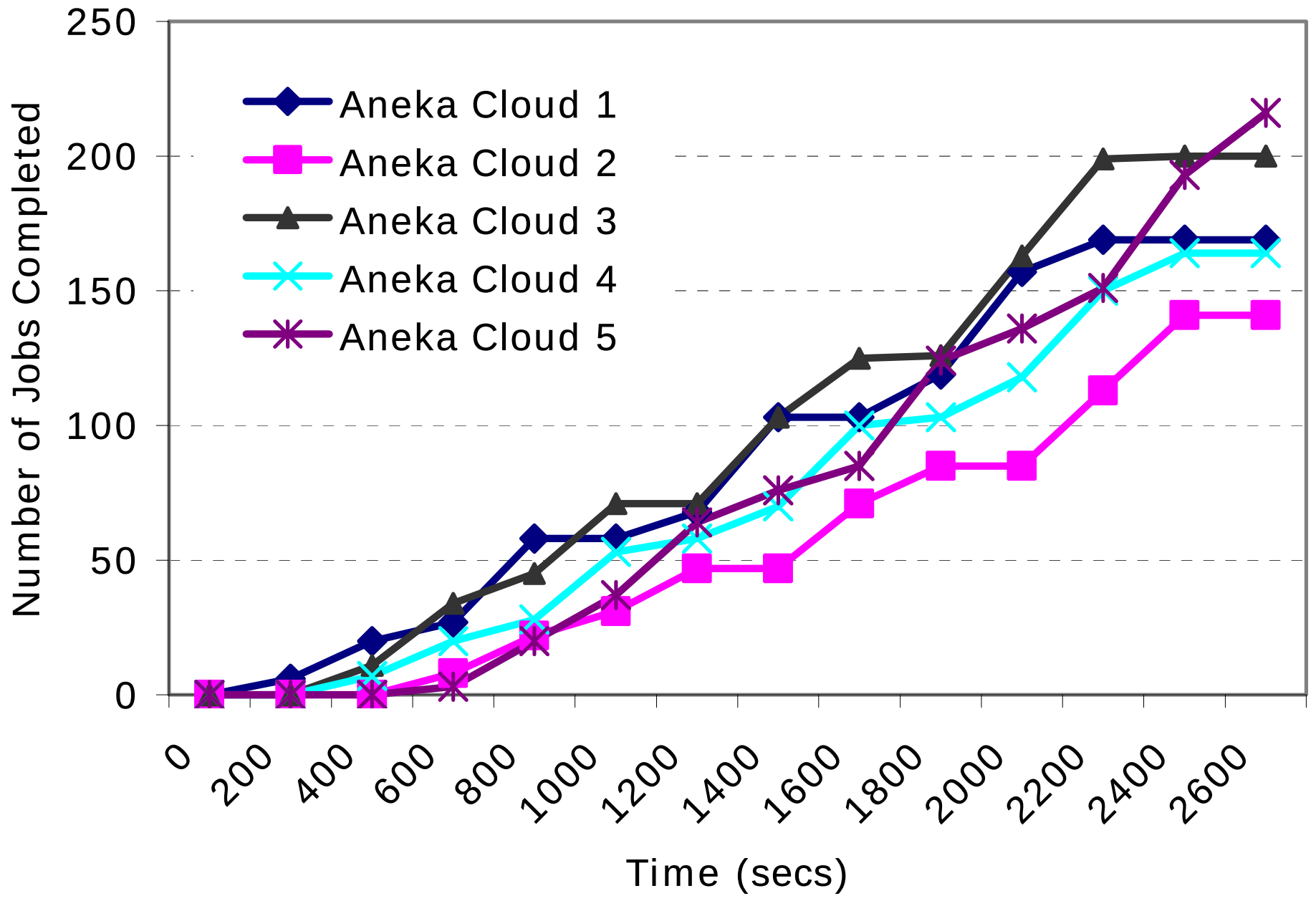

Figure 15: P2PThreadExecutionService: Time (secs) vs. Number of Jobs Completed.

Figure 16 shows the aggregate percentage of task and thread jobs processed by the nodes of the different Aneka Clouds in the federation. As mentioned in our previous discussions, Aneka Clouds 3, 4, and 5 ends up processing larger percentage for both Task and Thread application composition models. Together they process approximately 140% of total 200% jobs (100% task + 100% thread) in the federation.

## 7. Related Work

Volunteer computing systems including Seti@home [1] and Boinc [10] are the first generation implementation of public resource computing systems. These systems are engineered on the traditional master/worker model, wherein a centralized scheduler/coordinator is responsible for scheduling, dispatching tasks and collecting data from the participant nodes in the Internet. These systems do not provide any support for multi-application and programming models, a capability which is inherited from the Aneka to the Aneka-Federation platform. Unlike Seti@home and Boinc, Aneka-Federation creates a decentralized overlay of Aneka Enterprise Clouds. Further, Aneka-Federation allows submission, scheduling, and dispatching of application from any Aneka-Coordinator service in the system, thus giving every enterprise Cloud site autonomy and flexibility as regards to decision making.

OurGrid [11] is a peer-to-peer middleware infrastructure for creating an Internet-wide enterprise Grid computing platform. The message routing and communication between the OurGrid sites is done via broadcast messaging primitive based on the JXTA [12] substrate. ShareGrid [13] project extends the OurGrid infrastructure with fault-tolerance scheduling capability by replication tasks across a set of available nodes. In contrast to the OurGrid and the ShareGrid, Aneka-Federation implements a coordinated scheduling protocol by embedding a d-dimensional index over a DHT overlay, which makes the system highly scalable and guarantees deterministic search behavior (unlike JXTA). Further, the OurGrid system supports only the parameter sweep application programming model, while the Aneka-Federation supports more general programming abstractions including Thread, Task, and Dataflow.

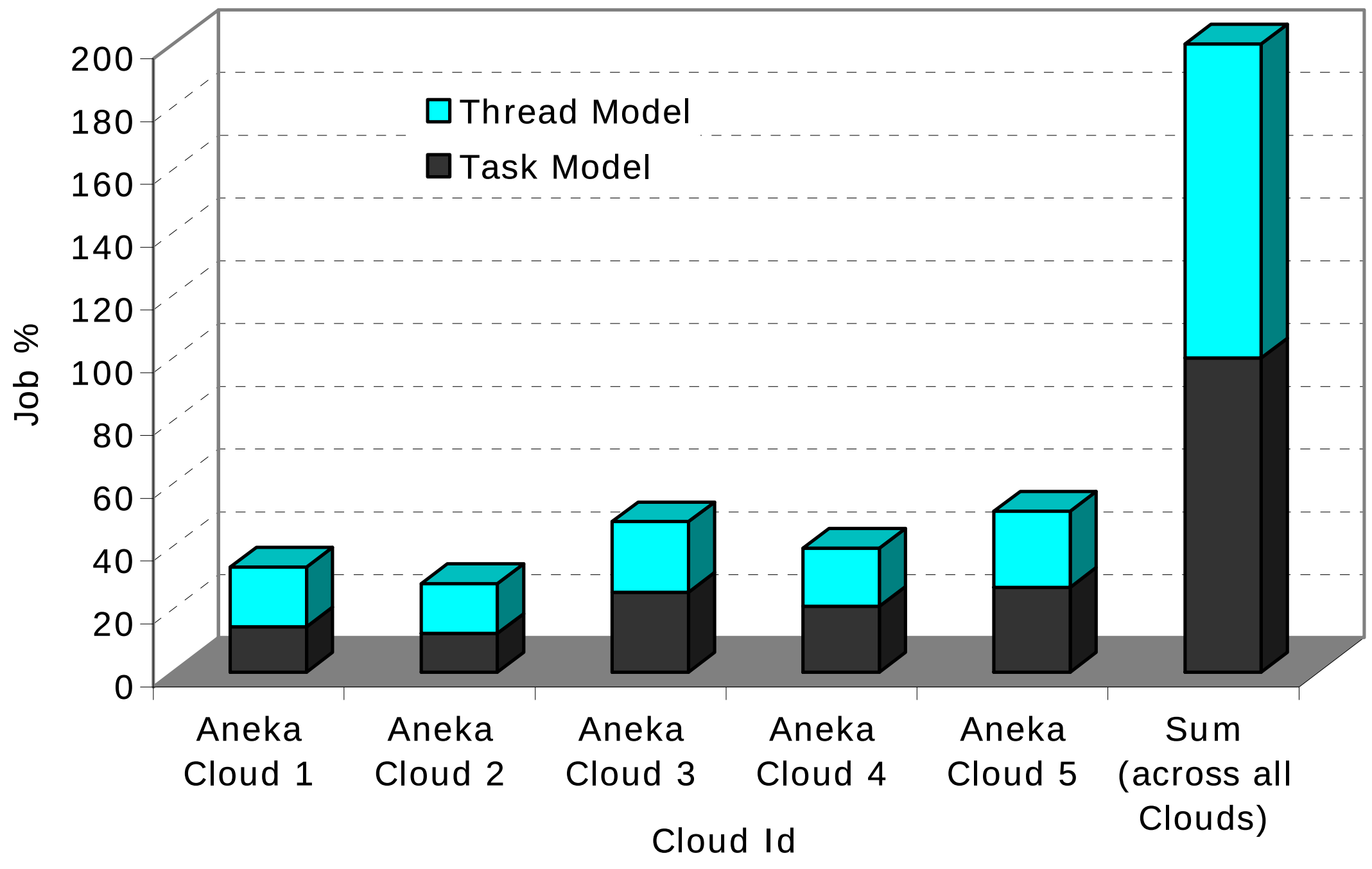

Figure 16: Enterprise Cloud Id vs. Job %.

Peer-to-Peer Condor flock system [14] aggregates Internet-wide distributed condor work pools based on the Pastry overlay [15]. The site managers in the Pastry overlay accomplish the load-management by announcing their available resources to all sites, who's Identifiers (IDs) appear in the routing table. An optimized version of this protocol proposes recursively propagating the load-information to the sites who's IDs are indexed by the contacted site's routing table. The scheduling coordination in an overlay is based on probing each site in routing table for resource availability. The probe message propagates recursively in the network until a suitable node is located. In the worst case, the number of messages generated due to recursive propagation can result into broadcast communication. In contrast, Aneka-Federation implements more scalable, deterministic and flexible coordination protocol by embedding a logical d-dimensional index over DHT overlay. The *d*-dimensional index gives the Aneka-Federation the ability to perform deterministic search for Aneka services, which are defined based on the complex node attributes (CPU type, speed, service type, utilization).

XtermWeb-CH [16] extends the XtermWeb [17] project with the functionalities such as peer-to-peer communication among the worker nodes. However, the core scheduling and management component in XtermWeb-CH, which is called the coordinator, is a centralized service that has a limited scalability. G2-P2P [18] uses the Pastry framework to create a scalable cycle-stealing framework. The mappings of objects to nodes are done via Pastry routing method. However, the G2-P2P system does not implement any specific scheduling or load-balancing algorithm that can take into account the current application load on the nodes and based on that perform runt-time load-balancing. In contrast, the Aneka-Federation realizes a truly decentralized, cooperative and coordinated application scheduling service that can dynamically allocate applications to the Aneka services/nodes without over-provisioning them.

## 8. Conclusion and Future Directions

The functionality exposed by the Aneka-Federation system is very powerful, and our experimental results on real test-bed prove that it is a viable technology for federating high throughput Aneka Enterprise Cloud systems. One of our immediate goals is to support substantially larger Aneka-Federation setups than the ones used in the performance evaluations. We intend to provide support for composing more complex application models such as e-Research workflows that have both compute and data node requirement. The resulting Aneka-Federation infrastructure will enable new generation of application composition environment where the application components, Enterprise Clouds, services, and data would interact as peers.

There are several important aspects of this system that require further implementation and future research efforts. One such aspect being developing fault-tolerant (self-healing) application scheduling algorithms that can ensure robust executions in the event of concurrent failures and rapid join/leave operations of enterprise Clouds/Cloud nodes in decentralized Aneka-Federation overlay. Other important design aspect that we would like to improve is ensuring a truly secure (self-protected) Aneka-Federation infrastructure based on peer-to-peer reputation and accountability models.

## 9. Acknowledgements

The authors would like to thank Australian Research Council (ARC) and the Department of Innovation, Industry, Science, and Research (DIISR) for supporting this research through the Discovery Project and International Science Linkage grants respectively. We would also like to thank Dr. Tejal Shah, Dr. Sungjin Choi, Dr. Christian Vecchiola, and Dr. Alexandre di Costanzo for proofreading the initial draft of this chapter. The chapter is partially derived from our previous publications [3].